\documentclass[10pt, conference, compsocconf]{IEEEtran}

\usepackage{amsmath, amsthm, amssymb}
\usepackage{amssymb,epsfig}

\ifCLASSINFOpdf
\else
\fi

\newtheorem{Lemma}{Lemma}

\newtheorem{Theorem}[Lemma]{Theorem}
\newtheorem{Proposition}[Lemma]{Proposition}
\newtheorem{Corollary}[Lemma]{Corollary}
\newtheorem{Definition}[Lemma]{Definition}

\newcommand{\imp}{\mathop{\mathrm{imp}}\nolimits}
\newcommand{\row}{\mathop{\mathrm{row}}\nolimits}
\newcommand{\clique}{\mathop{\mathrm{clique}}\nolimits}
\newcommand{\degree}{\mathop{\mathrm{degree}}\nolimits}
\newcommand{\mixed}{\mathop{\mathrm{mixed}}\nolimits}
\def\qed{\hskip 3pt \hbox{\vrule width4pt depth2pt height6pt}}

\begin{document}
%
\title{On some sufficient conditions for distributed Quality-of-Service support in wireless networks}


\author{\IEEEauthorblockN{Ashwin Ganesan}
\IEEEauthorblockA{(Formerly with ECE dept, University of Wisconsin at Madison, USA)\\
Current Address: 53 Deonar House\\
Deonar Village Road, Mumbai-88, India\\
Email: ashwin.ganesan@rediffmail.com}
}


%


\maketitle

\begin{abstract}
Given a wireless network where some pairs of communication links interfere with each other, we study sufficient conditions for determining whether a given set of minimum bandwidth Quality of Service (QoS) requirements can be satisfied.  We are especially interested in algorithms which have low communication overhead and low processing complexity.  The interference in the network is modeled using a conflict graph whose vertices are the communication links in the network.  Two links are adjacent in this graph if and only if they interfere with each other due to being in the same vicinity and hence cannot be simultaneously active.  The problem of scheduling the transmission of the various links is then essentially a fractional, weighted vertex coloring problem, for which upper bounds on the fractional chromatic number are sought using only localized information.  We present some distributed algorithms for this problem, and discuss their worst-case performance.   These algorithms are seen to be within a bounded factor away from optimal for some well known classes of networks and interference models.
\end{abstract}


%
\IEEEpeerreviewmaketitle

\section{Introduction}

In recent years there has been an increasing interest in using data networks to support a wide variety of applications, each requiring a different Quality of Service (QoS).  For example, real-time applications such as voice, video and industrial control are time-sensitive and require that the delay be small, while for other data applications the sender may require that a constant, minimum bit-rate service be provided.  In the simplest and lowest level of service, such as the one provided in the Internet Protocol service model, the network makes a best-effort to deliver data from the source to destination, but it makes no guarantees of any kind, so it is possible that packets can get dropped, delayed, or delivered out of order. However, this basic level of service is insufficient for many data applications such as videoconferencing that also have a minimum bandwidth requirement.  We consider in this work applications requiring a minimum bandwidth quality-of-service.

Consider a wireless communication network where nodes (which represent wireless devices such as laptops, phonoes, routers, sensors, etc) wish to communicate with each other using a shared wireless medium.  Any given pair of nodes may make a request for a dedicated point-to-point link between them that supports their required bit-rate Quality of Service.  The objective of the the \emph{admission control} mechanism is to decide whether the desired service can be provided, given the available resources, without disrupting the service guaranteed to previously admitted demands.  This mechanism needs to take into account the fact that nodes in the same vicinity contend for the shared wireless medium and hence can cause interference if simultaneously active. Also, for reasons such as low communication overhead and scalability, it is desired that this decision be made in some decentralized fashion.  Thus, each node may have access only to information pertaining to its local neighborhood and not about the entire communication network.  These two requirements - that the decision take into account \emph{interference} due to neighboring nodes and that it be made in a \emph{decentralized} manner - are crucial aspects of the particular problem we study.  If a decision to admit a demand is made, the \emph{scheduling} problem is to schedule the transmissions of the various nodes so as to provide the service level that was guaranteed.  The focus of most of our work here is on the admission control problem and not the scheduling problem.  For an introduction to the flow admission control problem, see \cite{Bertsekas:Gallager:1992}.

More formally, the wireless network model and desired QoS are specified as follows.
Let V be a set of nodes and $L \subseteq V \times V$ be a set of
communication links.   Each link $(i,j) \in L$ makes a demand (to
transmit information from $i$ to $j$) at a rate of $f(u,v)$ b/s.  The
total bandwidth of the shared wireless medium available for the communication network $G=(V,L)$ is $C$~b/s. The main problem studied here is to determine whether a set of demands $(f(\ell):\ell \in
L)$ can be satisfied.  Of course, if all the links can be simultaneously
active, the set of demands $(f(\ell): \ell \in L)$ can be satisfied as long
as each individual demand is at most $C$.  However, due to
interference effects nodes in the same vicinity contend for the shared
wireless medium and hence cannot be active at the same time.
For example, in IEEE 802.11 MAC
protocol-based networks, any nodes adjacent to node $i$ or to node
$j$ are required to be idle while the communication $(i,j)$ takes
place. 

The interference in the network is modeled using a \emph{conflict graph}.  Given a network graph
$G = (V,L)$, define its conflict graph to be $G_C = (L,L')$, where
two links $\ell_1$ and $\ell_2$ are adjacent in the conflict graph if and
only if they cannot be simultaneously active.  The conflict graph $G_C=(L,L')$ specifies which pairs of links
interfere with each other.  This interference
model has been studied recently by a number of authors; for example,
see Jain et al \cite{Jain:Padhye:etal:03}, Hamdaoui and Ramanathan
\cite{Hamdaoui:Ramanathan:05}, and  Gupta, Musacchio and Walrand
\cite{Gupta:Musacchio:Walrand:07}.  A special case of this model,
where two links are considered to be interfering if and only if they
are incident (in the network graph) to a common node, has been studied earlier by Hajek
\cite{Hajek:1984}, Hajek and Sasaki \cite{Hajek:Sasaki:1988}, and
Kodialam and Nadagopal \cite{Kodialam:Nandagopal:2005}.   

In the admission control problem studied in this work, the vertices of the conflict graph correspond to links in the communication network.  The quality-of-service metric is specified in terms of the bandwidth desired by each link.  This gives rise to one demand value for each vertex of the conflict graph, and this value could possibly be non-integral.  The admission control problem is then to determine whether these demands can be satisfied using a specified amount of resource (total available bandwidth).   This problem is different from the classical weighted vertex coloring problem in some ways.  First, fractional solutions to the coloring problem are also admissible, as indicated by the linear programming formulation given below. Second, our emphasis is on decisions that can be made in a decentralized manner, i.e. using only localized information.  Finally, the conflict graph sometimes has additional structure derived from the structure of the links in the network or the interference model.   The admission control problem studied here is essentially that of obtaining, using only localized information, an upper bound on the resource required to satisfy a demand pattern. The scheduling problem, which we do not study here, concerns how these resources are actually allocated or managed.

This paper is organized as follows.  The admission control problem studied here is formulated precisely in Section~\ref{sec:model:problem:formulation}.  Decentralized solutions to this problem are then studied:  Section~\ref{sec:row:constraints} is on the row constraints, Section~\ref{sec:degree:mixed} is on the degree and mixed conditions, and Section~\ref{sec:clique:constraints} is on the clique constraints.  These results provide sufficient conditions and distributed algorithms for admission control.  The main new results here concern the worst-case performance of these algorithms.   Finally, in Section~\ref{section:examples} the results obtained thus far are applied to some specific examples such as networks with primary interference constraints.  

\subsection{Model and problem formulation}
\label{sec:model:problem:formulation}
We first state the flow control problem
formally. Then, in order to avoid repeating trivialities throughout the
paper, we will present an equivalent reformulation of the problem that
ignores many of the
constants and variables and involves just the essential details. We
will work only with this reformulation in the rest of this paper.

Let $G=(V,L)$ be a network graph, where $L \subseteq V \times V$.
Each link $\ell \in L$ has a maximum transmission capacity of $C_\ell \le C$
b/s, and there is a demand to use that link at some rate $f(\ell)$ b/s; the
total available bandwidth of the shared wireless medium is $C$ b/s.
The main problem we study is to determine, using only localized information, whether the set of demands $(f(\ell): \ell \in L)$
can be satisfied.  Setting $\tau(\ell) = f(\ell)/C_{\ell}$, we get an equivalent reformulation where each link makes a demand to be active for a certain fraction of every unit of time.  

More precisely, an independent set of a graph
$G_C=(L,L')$ is a subset $I \subseteq L$ of elements that are pairwise nonadjacent.  If the set of links that are simultaneously active is an independent set, then these links cause no interference with each other and can have (a part of) their demands satisfied during the same time slot.
 Let  $\mathcal{I}(G_C)$ denote the set
of all independent sets of $G_C$. 
 A \emph{schedule} is a map $t:\mathcal{I}(G_c) \rightarrow
 \mathbb{R}_{\ge 0}$.  The schedule assigns to each independent set
 $I_j$ a time duration $t_j = t(I_j)$, which specifies the fraction of
 time that  the links in $I_j$ are active.  A schedule $t$ is said to satisfy a set
 of demands $(f(\ell): \ell \in L)$ if, for each $\ell \in L$, $\sum_{I_j:\ell \in
   I_j} t_j C_l \ge f(\ell)$, and if the duration of the schedule
 $\sum_{I_j} t_j$ is at most 1.  A schedule is said to be optimal if
 it satisfies the demand of all the links and has minimum duration.


\bigskip \noindent \textbf{Reformulation.} Suppose we are given a network graph
$G=(V,L)$ and a conflict graph that specifies which links interfere
with each other. Let $\tau(\ell)$ denote the
amount of time when link $\ell$ demands to be
active. A link demand vector $(\tau(\ell): \ell \in L)$
 is said to be feasible within
 time duration $[0,T]$ if
 there exists a schedule of duration at most $T$ that satisfies the
 demands.  We will often assume, for simplicity of exposition, that
 $T=1$.
Note that a schedule is a map  $t:\mathcal{I}(G_c) \rightarrow
 \mathbb{R}_{\ge 0}$ that assigns to each independent set $I_j$ of the
 conflict graph a time
 duration $t(I_j)$. A link $\ell$ is then active for total duration
 $\sum_{j:\ell \in I_j} t(I_j)$.
The flow admission problem is to determine whether there exists a schedule of
duration at most $T$ that satisfies the link demand vector $\tau$.  The
scheduling problem is to realize such a schedule.  We are interested in solutions that can be implemented using only localized information and with low processing cost.

\bigskip \noindent \textbf{Notation.}
It will be convenient to use the following notation.  Let $G=(V,E)$ be a simple, undirected graph. For $v \in V$,
$\Gamma(v)$ denotes the neighbors of $v$. $\alpha(G)$ denotes the
maximum number of vertices of $G$ that are pairwise nonadjacent. For
$V' \subseteq V$, $G[V']$ denotes the induced subgraph whose
vertex-set is $V'$ and whose edge-set is those edges of $G$ that
have both endpoints in $V'$. For any $\tau: V \rightarrow \mathbb{R}$
and any $W \subseteq V$, define $\tau(W):= \sum_{v  \in W}
\tau(v)$.

\subsection{Prior work and our contributions}

The take off point for our work is the prior work of \cite{Gupta:Musacchio:Walrand:07} and  \cite{Hamdaoui:Ramanathan:05}.  Their work proves that certain distributed algorithms provide sufficient conditions for admission control.  They call these conditions the row constraints \cite{Gupta:Musacchio:Walrand:07} (or rate condition \cite{Hamdaoui:Ramanathan:05}), the degree condition \cite{Hamdaoui:Ramanathan:05}, mixed condition \cite{Hamdaoui:Ramanathan:05}, and scaled clique constraints \cite{Gupta:Musacchio:Walrand:07}.

The main results of this paper are along the following lines: the exact worst-case performance of these distributed admission control mechanisms is characterized and it is thereby shown that these mechanisms can be arbitrarily far away from optimal; we then show that for some well known classes of networks and interference models, these distributed algorithms are actually within a bounded factor away from optimal.

\section{Row constraints}
\label{sec:row:constraints}
We now present a sufficient condition for flow admission control that can be implemented distributedly. Given a conflict graph $G_C=(L,L')$, link
demand vector $(\tau(\ell):\ell \in L)$ and $T$, a sufficient condition for feasibility is given by the following result  (see \cite[Thm.~1]{Hamdaoui:Ramanathan:05}, \cite[Thm.~1]{Gupta:Musacchio:Walrand:07}):

\begin{Proposition}
\label{prop:row:constraints:sufficient}
If $\tau(\ell) + \tau(\Gamma(\ell)) \le T$ for each $\ell \in L$, then the
demand vector $(\tau(\ell): \ell \in L)$ is feasible within duration $T$ .

\end{Proposition}


These constraints are called the row constraints in
\cite{Gupta:Musacchio:Walrand:07} and the rate condition in
\cite{Hamdaoui:Ramanathan:05}.  Let $A=[a_{ij}]$ be the 0-1 valued
$n \times n$ adjacency matrix of $G_C$ where $a_{ij} = 1$ iff $i=j$
or $\ell_i$ and $\ell_j$ are interfering.  Let $\mathbf{1}$ denote the vector whose every entry is 1.  Then the sufficient condition
above is equivalent to the condition $A \tau \le  T\mathbf{1}$ on the
rows of $A$, hence the name \emph{row constraints}.

The proof of Proposition~\ref{prop:row:constraints:sufficient} gives
a very efficient algorithm for checking feasibility. It provides
both a distributed admission control mechanism as well as a distributed
scheduling algorithm: when a link $\ell_i$ that is currently inactive
makes a demand to be active for duration $\tau(\ell_i)$, the admission
control mechanism can be implemented efficiently by just checking
the condition above for $\ell_i$ and its neighbors.  The
information required by a link to check this condition is just its
demand and the demand of its neighbors.  Furthermore, the distributed scheduling algorithm that meets the demand for link $\ell_i$  needs to know only the time intervals already
assigned to the neighbors of $\ell_i$ in order to determine the time
interval for $\ell_i$.


\subsection{Row constraint polytope and induced star number}

Given a
conflict graph, let $P_I$ denote its independent set polytope. This
polytope is defined as the convex hull of the characteristic
vectors of the independent sets of the graph.  Note that $P_I$ is exactly equal
to the set of all link demand vectors which are feasible within one
unit of time.  For the given conflict graph, let $P_{\row}$ denote the set of all link demand
vectors that satisfy the row constraints for $T=1$; that is,
$$P_{\row} := \{ \tau \ge 0: \tau(\ell_i) + \tau(\Gamma(\ell_i)) \le 1 ~\forall i\}.$$
Since the row constraints are sufficient, $P_{\row} \subseteq P_I$.
Also, note that $\beta P_{\row} = \{ \tau:
\tau(\ell_i)+\tau(\Gamma(\ell_i)) \le \beta\}$.  Define the scaling
factor $$\beta_{\row} := \inf\{\beta \ge 1: P_I
\subseteq \beta_{\row} P_{\row}  \}.
$$
Equivalently, $$\beta_{\row} = \sup_{\tau \in P_I} \max_i \{
\tau(\ell_i) + \tau(\Gamma(\ell_i))\}.
$$
So $P_{\row} \subseteq P_I \subseteq \beta_{\row}
P_{\row}$, and $\beta_{\row}$ is the smallest scaling factor which converts the sufficient condition into a necessary one.

It has been pointed out (cf. \cite{Gupta:Musacchio:Walrand:07}) that the row constraints can be
arbitrarily far away from optimal. For example, suppose the network
consists of links $\ell_1,\ldots,\ell_{d+1}$, where $\ell_1$ interferes with
each of $\ell_2,\ldots,\ell_{d+1}$ and there is no interference between the remaining links.
Then the conflict graph is a star graph. The link demand vector
$(\varepsilon,1-\varepsilon,\ldots,1-\varepsilon)$ is feasible
within one unit of time, but the row constraint for $\ell_1$ has value
$\tau(\ell_1)+\tau(\Gamma(\ell_1)) = \varepsilon + (1-\varepsilon)d$ which can be
made arbitrarily close to $d$ as $\varepsilon$ approaches 0.  This shows that $\beta_{\row}
\ge d$.  We prove next that the opposite inequality also holds, i.e. the row constraints can be a factor $s$
away from the optimal schedule time for some demand vector
only if the conflict graph contains a star on $\lceil s+1 \rceil$
vertices as an induced subgraph.

\begin{Definition}  The \emph{induced star number} of a graph $H$ is
defined by
$$ \sigma(H) := \max_{v \in V(H)} \alpha(H[\Gamma(v)]).\quad\quad\qed$$

\end{Definition}

Hence, the induced star number of a graph is the number of leaf
vertices in the maximum sized star of the graph.  This number determines
exactly how close the row constraints are to optimal in the worst case:
\begin{Theorem}
\label{theorem:star:number:row:factor}
 Let $G_C$ be a conflict graph.
The exact worst-case performance of the row constraints is given by $\beta_{\row} = \sigma(G_C)$.
\end{Theorem}

\noindent \emph{Proof:}  See \cite{Ganesan:2008}.
\hfill\qed

It follows that the row
constraints, which are sufficient conditions, are also necessary iff
$P_{\row} = P_I$, which is the case iff $\sigma(G_C)=1$, which is the case iff each
component of $G_C$ is a complete graph.  Thus, the row constraints above are also a necessary condition if and only if the conflict graph is the disjoint union of complete graphs.   While the induced star number of a graph can be arbitrarily large, for special classes of networks studied in the literature this quantity is bounded by a fixed constant. This happens to be in the case for unit disk graphs and for networks with primary interference constraints.

\subsection{A strengthening of the row constraints}

We showed above that the performance of the row constraints is determined by the induced star number $\sigma(G_C)$. We now show that a slight improvement to $\sigma(G_C)-1$ can be obtained.  For simplicity of exposition, we shall assume in this section that $G_C$ is connected; if this is not the case we can work with each connected component separately and the results here still apply.

Recall that the row constraint corresponding to link $\ell_i$ is that the sum total of the demand $\tau(\ell_i)$ and the demands of \emph{all} its interfering neighbors $\tau(\Gamma(\ell_i))$ not exceed the available resource $T$.  It is easy to see that all the links in the network, except for any one designated link, say $\ell_1$, can ignore the demand of up to \emph{one} of its interfering neighbors.

\begin{Proposition}
Given a network and its conflict graph $G_C$, pick any designated link $\ell_1 \in L$.  A sufficient condition for $\tau$ to be feasible within duration $T$ is that
\begin{eqnarray*}
\tau(\ell_i) + \tau(\Gamma(\ell_i))&&\le T,~ i=1 \\
\tau(\ell_i) + \{\tau(\Gamma(\ell_i)) - \min_{\ell_j \in \Gamma(\ell_i)} \tau(\ell_j) \}&&\le T, ~i=2,\ldots,m.
\end{eqnarray*}
\end{Proposition}

\noindent \emph{Proof:}  See \cite{Ganesan:2008}.
\hfill\qed

Note that this sufficient condition is equivalent to the row constraints when $G_C$ is complete.  
The following stronger result can be obtained when the graph is not an odd cycle.

\begin{Proposition}
Suppose $G_C$ is not complete.  Then the set of constraints
$$\tau(\ell_i) + \{\tau(\Gamma(\ell_i)) - \min_{\ell_j \in \Gamma(\ell_i)} \tau(\ell_j) \} ~\le ~T,~~~i=1,\ldots,m $$
is a sufficient condition for $\tau$ to be feasible within duration $T$ if $G_C$ is not an odd cycle.
Furthermore, the smallest scaling factor that converts this sufficient condition into a necessary one is equal to exactly $\sigma(G_C)$ or $\sigma(G_C)-1$, depending on the structure of $G_C$.
\end{Proposition}

\noindent \emph{Proof:} See \cite{Ganesan:2008}.
\hfill\qed

\section{Degree and mixed conditions}
\label{sec:degree:mixed}
It was shown that the row constraints provided a simple, distributed sufficient condition for feasibility of a given demand vector. In this condition, there was exactly one constraint associated with each link, namely, the sum total of the demand of the link and demands of its neighbors not exceed the available resource. We now describe an even simpler condition.  We call this the degree condition since it requires knowing, for each link, the demand of that link and just the \emph{number} (not actual demands) of links interfering with it.

Suppose link $\ell_i$ interferes with exactly $d(\ell_i)$ other links, i.e. in the conflict graph $\ell_i$ has degree $d(\ell_i)$.  Then, the following result provides another sufficient condition for admission control \cite{Hamdaoui:Ramanathan:05}:
\begin{Proposition}
\label{proposition:degree:condition:sufficient}
A given demand vector $(\tau(\ell_i): \ell_i \in L)$ is feasible within duration $T$ if $\tau(\ell_i) (d(\ell_i)+1) \le T$ for each $\ell_i \in L$.
\end{Proposition}

The performance of the degree condition is determined, not surprisingly, by the maximum degree of a vertex in the conflict graph.  More precisely, define
$$P_{\degree} := \{\tau \ge 0: \tau(\ell_i)(d(\ell_i)+1) \le 1,~ \forall i\}.$$
Then $P_{\degree} \subseteq P_I$ by Proposition~\ref{proposition:degree:condition:sufficient}.   Define $$\beta_{\degree} := \inf\{\beta \ge 1: P_I \subseteq \beta P_{\degree} \}.$$  Let $\Delta(G_C)$ denote the maximum degree of a vertex in $G_C$.

\begin{Lemma}
For any conflict graph $G_C$,  the exact worst-case performance of the degree condition is given by  $\beta_{\degree}(G_C) = \Delta(G_C)+1$.
\end{Lemma}

\noindent \emph{Proof:} See \cite{Ganesan:2008}.
%
\hfill\qed

This implies that the sufficient degree condition is also necessary (and hence optimal) iff $G_C$ is the empty graph, i.e. iff no two links interfere with each other.  It is possible to combine the row constraints and degree constraints to get a sufficient condition which is strictly stronger, as shown in \cite{Hamdaoui:Ramanathan:05}:

\begin{Proposition}
A link demand vector $\tau$ is feasible within duration $T$ if
$$ \min \{~\tau(\ell_i)+\tau(\Gamma(\ell_i))~,~\tau(\ell_i)(d(\ell_i)+1) ~\} \le T, ~\forall \ell_i \in L.$$
\end{Proposition}

In general,
$$P_{\row},P_{\degree}~ \varsubsetneq ~(P_{\row} ~\cup~ P_{\degree}) ~\varsubsetneq ~P_{\mixed} ~\varsubsetneq~ P_I,$$
where
$$P_{\mixed} := \{\tau: \min \{\tau(\ell_i)+\tau(\Gamma(\ell_i)),$$ $$
\tau(\ell_i)(d(\ell_i)+1) \} \le 1,~\forall \ell_i \in L \}.$$

Let $\beta_{\mixed}$ denote the smallest scaling factor that converts the sufficient mixed condition into a necessary one; hence, given the conflict graph $G_C=(L,L')$ and its independent set polytope $P_I$, we have that $P_{\mixed} \subseteq P_I \subseteq \beta_{\mixed} P_{\mixed}$ and
$$ \beta_{\mixed} = \sup_{\tau \in P_I}~ \max_{\ell_i \in L}~ \min\{\tau(\ell_i)+\tau(\Gamma(\ell_i)),\tau(\ell_i)(d(\ell_i)+1)  \} .$$

\begin{Theorem}
\label{theorem:mixed:condition:performance}
The worst-case performance of the mixed condition is bounded as
$$ \frac{1+\sigma(G_C)}{2} ~\le ~ \beta_{\mixed} ~ \le ~ \sigma(G_C),$$
where $\sigma(G_C)$ denotes the induced star number of $G_C$.  Moreover, the lower and upper bounds are tight; the star graphs realize the lower bound, and there exist graph sequences for which $\beta_{\mixed}$ approaches the upper bound arbitrarily closely.
\end{Theorem}

\noindent \emph{Proof:} See \cite{Ganesan:2008}.
\hfill\qed

One general class of graphs that includes the star graphs, the even and odd cycles, the complete graphs and bipartite graphs are those that satisfy the following property: for each vertex $\ell \in L$ in the graph $G_C=(L,L')$, the neighbors of $\ell$ induce a disjoint union of complete graphs.  For this general class of graphs there is a simple expression for the exact value of $\beta_{\mixed}$:

\begin{Theorem}
\label{theorem:mixed:performance:exact}
Suppose $G_C=(L,L')$ satisfies $\sigma(G_C[\Gamma(\ell)]) \le 1, ~\forall \ell \in L$.   Let $d(\ell)$ denote the number of neighbors of $\ell$ and let $\eta_{\ell}$ denote the number of connected components induced by the neighbors of $\ell$.  Then
$$\beta_{\mixed} = \max_{\ell \in L} \frac{\eta_{\ell} (1+d(\ell))}{\eta_{\ell} + d(\ell)}.$$
\end{Theorem}


\section{Clique constraints}
\label{sec:clique:constraints}
A necessary condition for a given link demand vector to be feasible can be obtained as follows. Suppose there exists a schedule of duration 1 satisfying demand $\tau$.  Then if $K$ is a clique in the conflict graph, the time intervals assigned to the distinct links in $K$ must be disjoint, hence $\tau(K) \le 1$.  Thus, a necessary condition for $\tau$ to be feasible within duration $T$ is that $\tau(K) \le T$ for every maximal clique $K$ in the conflict graph. These constraints are called \emph{clique constraints} \cite{Gupta:Musacchio:Walrand:07}.  As before, we can associate a polytope with this necessary condition; define
$$P_{\clique} := \{\tau: \tau(K) \le 1, \forall  K \}  \supseteq P_I,$$  where $K$ runs over all the cliques (or equivalently, over just all the maximal cliques) of the conflict graph.

%
%
%
%

Using the notion of the imperfection ratio of graphs, bounds on the suboptimality of clique constraints were obtained \cite{Gupta:Musacchio:Walrand:07} for the case of unit disk graphs.  More precisely, given a conflict graph $G_C$ and demand vector $\tau$, let $T^*(\tau)$ denote the minimum duration of a schedule satisfying $\tau$ (the optimal value of this linear program is also the smallest $\beta$ such that $\tau \in \beta P_I$, and $T^*(\textbf{1})$ is often referred to as the fractional chromatic number of $G_C$). Let $T_{\clique}(\tau)$ denote the maximum value of $\tau(K)$ over all cliques $K$ in the conflict graph; so $T_{\clique}(\tau) \le T^*(\tau)$.  The imperfection ratio of a graph $G_C$ is defined as
$$\imp(G_C) := \sup_{\tau \ne 0} \{ T^*(\tau) / T_{\clique}(\tau) \} .$$
This quantity has been studied in \cite{Gerke:McDiarmid:2001}; it is finite and is achieved for any given graph.  In the definition above, for a given demand vector $\tau$, the numerator specifies the exact amount of resource required to satisfy the demand, as determined by an optimal, centralized algorithm.  The denominator specifies a lower bound on the resource required to satisfy the demand, as determined by a particular distributed algorithm (the clique constraints).  Their ratio is the factor by which the distributed algorithm is away from optimal for the given demand vector.  The imperfection ratio, which maximizes this ratio over all demand patterns, is then the worst-case performance of the distributed algorithm.

The following general result is implicit in \cite{Gupta:Musacchio:Walrand:07} (where the authors focus on unit disk graphs) and in \cite{Gerke:McDiarmid:2001}:

\begin{Proposition}
\label{prop:clique:constraints:imp}
The largest scaling factor which converts the necessary clique constraints into a sufficient condition is $1 / \imp(G_C)$; i.e. the worst-case performance of the clique constraints is given by
$$\sup\{\beta \le 1: \beta P_{\clique} \subseteq P_I \} = \frac{1}{\imp(G_C)};$$
and
$$ \frac{1}{\imp(G_C)} P_{\clique} \subseteq P_I \subseteq P_{\clique} .$$
\end{Proposition}

%

\section{Examples}
\label{section:examples}

In this section we apply the results obtained so far to some special classes of networks and interference models.   In Section~\ref{section:primary:interference}, we examine a model of interference called primary interference constraints, which has been well-studied in the literature for the centralized setting; we examine the distributed version of the problem here.

\subsection{Primary interference model}
\label{section:primary:interference}
Given a network $G=(V,L)$, suppose two links are considered to be interfering iff they share one or more endvertices in common.  We refer to this kind of interference as primary interference.   This interference model arises, for example, from the assumption that each node can communicate to only one other node at any given time.
This interference model is perhaps the most well studied in the literature; for example, see \cite{Hajek:1984}, \cite{Hajek:Sasaki:1988},  \cite{Kodialam:Nandagopal:2005}.  The conflict graph for such a network is called a \emph{line graph}.
 It can be shown that that if the conflict graph $G_C$ is a line graph then $\sigma(G_C) \le 2$.  It follows that for such networks the row constraints will be at most a factor 2 away from optimal.

More specifically, for this interference model, the row constraints on the conflict graph can be reformulated on the network graph $G=(V,L)$ as follows.  Suppose link $\ell_i$ is incident between nodes $u_i$ and $v_i$.  Given link demand vector $\tau$, let $\tau(u)$ denote the sum of the demands of all links incident in $G$ to node $u$. Then the row constraint $\tau(\ell_i) + \tau(\Gamma(\ell_i)) \le T$ in the conflict graph $G_C=(L,L')$ is equivalent to the constraint $\tau(u_i) + \tau(v_i) - \tau(\ell_i) \le T$ in the network graph. This equivalence yields the following sufficient condition:
\begin{Corollary}
\label{corollary:row:constraints:line:graph}
Let $G=(V,L)$ be a network graph, and suppose two links interfere if and only if they are incident to a common node.  Then $(\tau(\ell):\ell \in L)$ is feasible within duration $T$ if for each $\ell=\{u,v\}$, $\tau(u)+\tau(v)-\tau(\ell) \le T$.  This sufficient condition is a factor of at most 2 away from optimal. 
\end{Corollary}

Another distributed algorithm that can be used in networks having primary interference constraints is given by the clique constraints.  Trivially, a necessary condition for $\tau$ to be feasible within duration $T$ is that $\tau(K) \le T$ for all cliques $K$ in the conflict graph.  By Proposition~\ref{prop:clique:constraints:imp}, the performance of the clique constraints is determined by the imperfection ratio of the conflict graph.  It is known that the imperfection ratio of a line graph is at most 1.25 \cite[Prop.~3.8]{Gerke:McDiarmid:2001}.  This means that a sufficient condition for $\tau$ to be feasible wtihin duration $T$ is that $1.25 \tau(K) \le T$ for all cliques $K$ in the conflict graph.  Since $G_C$ is a line graph, each clique $K$ in $G_C$ corresponds either to a set of links $K \subseteq L$ that are all incident to a common node in the network graph $G=(V,L)$ or to a set of three links that form a triangle in the network graph.  For $v \in V$, let $\tau(v)$ denote the sum of the demands of all links incident to $v$ in the network graph.  

\begin{Theorem}
\label{theorem:clique:constraints:line:graph}
Let $G=(V,L)$ be a network graph, and suppose two links interfere if and only if they are incident to a common node. Then $(\tau(\ell): \ell \in L)$ is feasible within duration $T$ if $\tau(v) \le 0.8T,~\forall v \in V$ and $\tau(uv)+\tau(vw)+\tau(uw) \le 0.8T,~\forall u,v,w \in V$. This sufficient condition is a factor of at most 1.25 away from optimal.
\end{Theorem}

An important aspect of this result is that, though the number of maximal cliques in a general graph can grow exponentially with the size of the graph, the number of maximal cliques in a line graph grows only polynomially in the size of the graph.  
Thus, unlike for general graphs, for line graphs the clique constraints provide an efficient distributed algorithm for checking feasibility of a given demand vector.

\bigskip \noindent \emph{Remark}:  A result due to Shannon on the edge-coloring of multigraphs \cite{Shannon:1949} implies that: for a given network graph $G=(V,L)$, a sufficient condition for $(\tau(\ell): \ell \in L)$ to be feasible within duration $T$ is that $\tau(v) \le 2T/3,~\forall v \in V$.  Theorem~\ref{theorem:clique:constraints:line:graph} improves this bound from a factor of 2/3 to 0.8.  This improvement is possible because, unlike in the classical edge-coloring problem, fractional coloring solutions are also admissible in our framework.  Furthermore, the sufficient condition in Theorem~\ref{theorem:clique:constraints:line:graph} is less localized, in that each node in the network graph needs to know not only the sum total of the demands of all links incident to it, but also the demands of all links between its neighbors.  




\section{Concluding remarks}

We introduced the notion of the induced star number of a graph and show that it determines the exact worse-case performance of the row constraints.
The performance of two other sufficient conditions - namely the degree condition and mixed condition - was also studied.   Finally, the results obtained thus far are applied to some specific classes of networks and interference models.

 These results imply that for some special classes of networks and interference models, there exist simple, efficient, distributed admission control mechanisms and scheduling mechanisms whose performance is within a bounded factor away from that of an optimal, centralized mechanism.

 A detailed report containing proofs of the results presented here is available from the author \cite{Ganesan:2008}.
 
%
%
%
%
%
%
%
%
%


\section*{Acknowledgements}
Thanks are due to professor Parmesh Ramanathan for
 suggesting this direction of scaling the sufficient conditions.

 {
\bibliographystyle{plain}
\bibliography{winet}
}

\end{document}